\documentclass[aps,twocolumn,superscriptaddress]{revtex4-2}
\pdfoutput=1
\usepackage{graphicx}
\usepackage{dcolumn}
\usepackage{bm}
\usepackage{color}
\usepackage{amsmath}

\definecolor{nblue}{RGB}{28,130,185}
\definecolor{cgreen}{RGB}{76,153,0}
\definecolor{myorange}{RGB}{245,156,74}

\usepackage{hyperref}
\hypersetup{
  colorlinks=true,
  citecolor=cyan,%magenta,
  urlcolor=cgreen
}

\def \beq{\begin{eqnarray}}
\def \eeq{\end{eqnarray}}
\def \mo{(\mbox{\footnotesize{1}})}
\def \mt{(\mbox{\footnotesize{2}})}

\begin{document}

\title{Phase coexistence in the Non-reciprocal Cahn-Hilliard model}
\author{Suropriya Saha}
\email{suropriya.saha@ds.mpg.de}
\affiliation{Max Planck Institute for Dynamics and Self-Organization (MPIDS), D-37077 G\"ottingen, Germany}% Am Fa{\ss}berg 17,

\date{\today}

\begin{abstract}
We establish the criterion for the phase coexistence in a mixture of nonreciprocally interacting scalar densities.  For an arbitrary number of components the active pressure exists for a specific class of interactions, and when the free energy receives no contribution from cross couplings between spatial gradients of two different species. In this case, the pressure can be used to determine phase equilibrium, i.e. to construct binodals, and the active mixture can be mapped to a passive system with an effective free energy. For general interfacial tension, the pressure changes discontinuously across a flat interface which assumes the form of an active Laplace pressure in two dimensions.

\end{abstract}
%\keywords{}
\maketitle
{\it{Introduction}}  -- The principles governing phase equilibrium form the cornerstone of the physics of passive mixtures. The concepts first explored by Gibbs~\cite{Gibbs1948,Sollich2001} are relevant to this day several decades later in multiple areas of research, from membrane-less organelles~\cite{AnnuRevLLPS-Christoph-frank-14, Michael_Rosenreview_17, Niebel2019} to active phase separation~\cite{Wittkowski2014,Bubbly_PhysRevX.8.031080, Jaime2019, TailleurTactic_PRL, Dinelli2023, Matthew-PRL-2022}. The concept of pressure requires careful consideration when applied to active matter~\cite{Gompper2020,Marchetti_RevModPhys.85.1143} which consists of units that are driven at the microscopic scale and collectively operate away from equilibrium in a quantifiable~\cite{FodorWijland_2016, PrawarDadhichi2018, Nardini_PhysRevX.7.021007} manner. The role of an active pressure depends on details of the active driving-- recent works have argued the relevance \cite{GeneralisedThermo} or otherwise \cite{Solon2015} of a pressure like quantity in different incarnations of activity. 

\begin{figure}
	\centering
	\includegraphics[width= 0.8\linewidth]{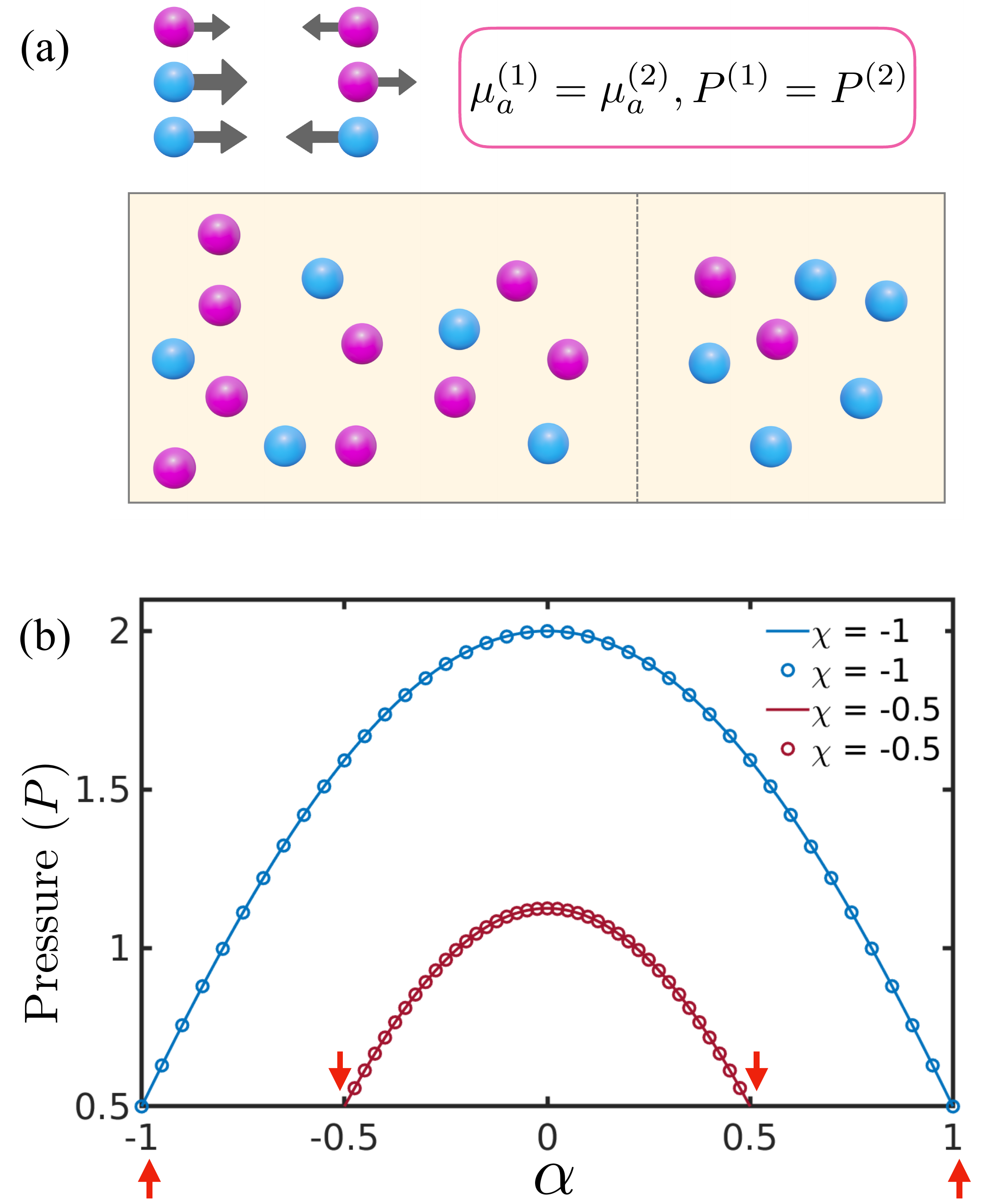}
	\caption{{\textbf{Pressure in the non-reciprocal Cahn-Hilliard model}} (a) Schematic showing the compartmentalisation of a system of multiple non-reciprocally interacting species. We find a quantity that plays the same role as pressure in determining phase equilibrium. Panel (b) shows the analytic expression for the active pressure matches very well with simulations as a function of the non-reciprocal parameter $\alpha$. The red arrows indicate the exceptional points corresponding to $|\alpha|<|\chi| $.  }
	\label{fig:Schematic}
\end{figure}

In this paper we ask a similar question in a new class of active matter systems, those driven by effective non-reciprocal interactions~\cite{Soto_PhysRevLett.112.068301,saha2019pairing}. The theoretical model that we explore in this paper in the Non-Reciprocal Cahn-Hilliard model (NRCH)~\cite{sahaPRX_2020,You_Marchetti_2020}, that describes the evolution of a system of multiple interacting conserved densities. The dynamical behaviour of NRCH, which emerges naturally in scalar mixtures with more than one conservation law~\cite{Thiele2023}, has been shown to be varied, ranging from travelling waves \cite{sahaPRX_2020,You_Marchetti_2020}, moving lattices, spiral defects \cite{rana2023defect}, Turing instabilities~\cite{Thiele_PhysRevE.103.042602} to spatiotemporal chaos \cite{Effervescence}. In this work we explore the possibility that a condition similar to pressure exists for non-reciprocally interacting conserved scalar densities. We focus on steady states of NRCH that do not break PT symmetry in contrast to~\cite{Fruchart2021,sahaPRX_2020,You_Marchetti_2020}, rather they evolve to a state with microscopic or macroscopic phase separated states with zero flux at the steady state, see Fig.~\ref{fig:Schematic} (a). 

{\it{Model and main results --}} The NRCH model is represented by the set of $N$ equations $\partial_t \phi_a = \nabla^2 \mu_a$ for $N$ conserved scalar densities $\phi_a$, where $\mu_a$ is the chemical potential for the $a$-th species. The system of equations represents an active mixture as $\mu_a$ is not derivable from a free energy. For simplicity in calculations we assume a polynomial form for $\mu_a$ as follows
\beq 
\mu_a  = \frac{\mbox{d} f_a}{ \mbox{d}\phi_a} + \sum_{b \neq a} \chi_{ab}\frac{ \partial f_{ab}}{ \partial \phi_a} - \sum_b {K}_{ab} \nabla^2 \phi_b,
\label{eq:chemicalPotential}
\eeq
where $f_a$ is a function of $\phi_a$ only, and $f_{ab}$ is a function of $\phi_a$ and $\phi_b$, where $a$ and $b$ are distinct from one other. The coefficients of interaction $\chi_{ab}$ and the interfacial tension $K_{ab}$ are arbitrary constants. The system describes a passive mixture for $\chi_{ab} = \chi_{ba}$ and $K_{ab} = K_{ba}$. Asymmetric coefficients both in the bulk free energy and the interfacial tension arises naturally in systems with nonreciprocal interactions, for example in a collection of phoretic Janus colloids~\cite{tucci2024nonreciprocal} or mixtures of quorum sensing particles~\cite{BenoitYu2023}. For $f_{ab} = \phi_a \phi_b$, Eq. \eqref{eq:chemicalPotential}, reduces to the form explored in \cite{sahaPRX_2020}. Eq.~\eqref{eq:chemicalPotential} represents a system with pair wise interactions between the species, an exploration of the consequences with three species interactions permitting terms of the form $\phi_a \phi_b \phi_c$ for example, is beyond the scope of the present work.

 We present a summary of our results before elucidating them in detail. The interactions are chosen such that the system undergoes bulk phase separation where multiple compartments, see Fig~\ref{fig:Schematic} (a) are formed within which the concentrations of all species are uniform. The interfaces separating the compartments are of vanishing width with a length-scale determined by the interfacial tension. Within an interface the densities change rapidly from one stationary composition to another. Integrating the product of the chemical potential and the gradient of each species across a flat interface separating two or more bulk phases, and summing over all the $N$ integrals so obtained, we find a pressure like quantity $P$ that is constant throughout the system. $P$ is used to determine phase coexistence as exactly one does in passive systems. $P$ is related to a free energy that has exactly the same functional form as the Lyapunov cost function of the minimal oscillator associated with the dynamics~\cite{sahaPRX_2020}. The procedure can be extended to multi-component systems with multiple conserved densities by constraining the interaction parameters $\chi_{ab}$. $P$, however, changes discontinuously on introducing coupling between the gradients of the interacting densities in the free energy, i.e. for non vanishing $K_{ab}$, for $a \neq b$ implying that a Laplace pressure develops even at flat interfaces. 
 \begin{figure}
	\centering
	\includegraphics[width= 0.8\linewidth]{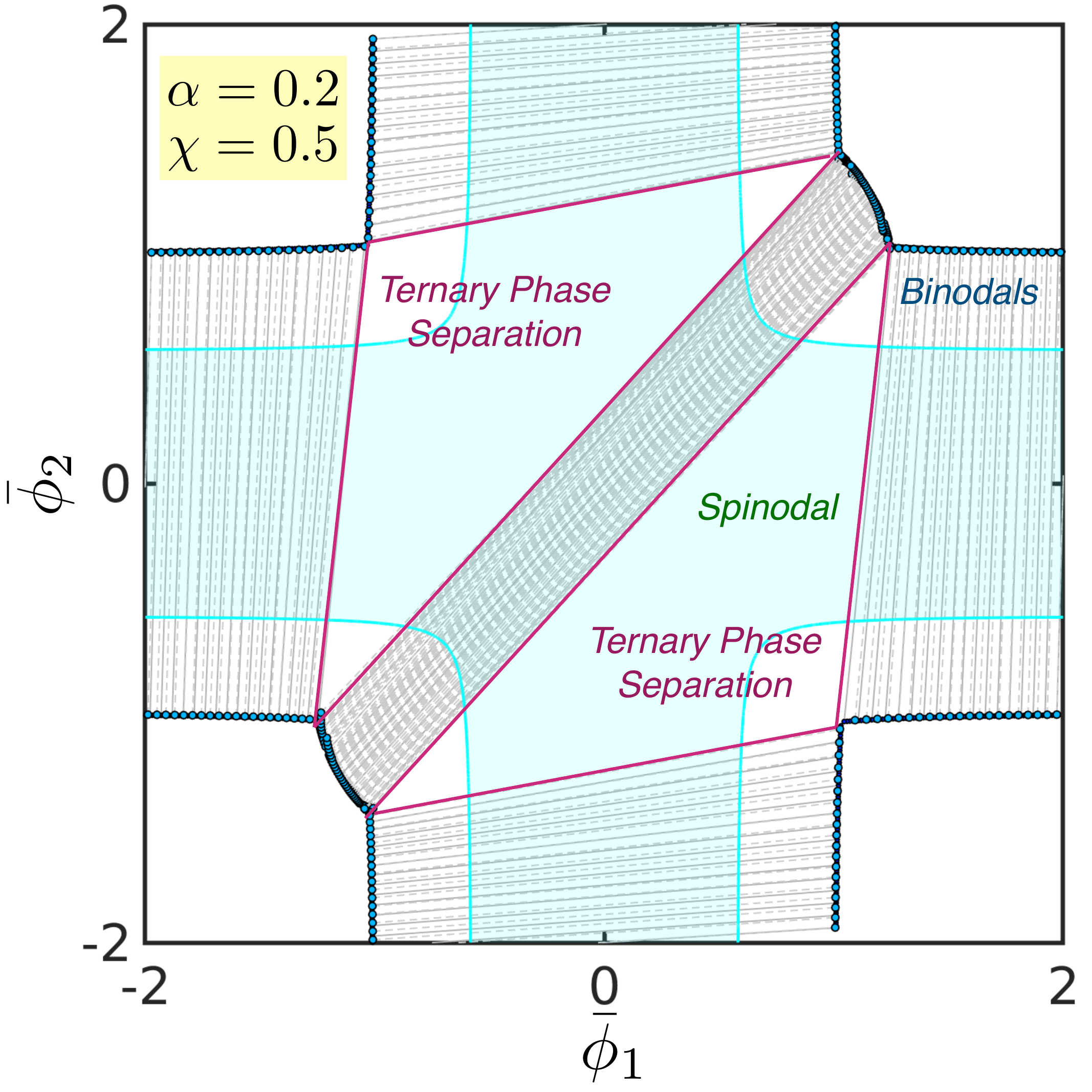}
	\caption{{\textbf{Phase coexistence:}} Figure showing a comparison of the binodals calculated using the conditions for equality of $P$ and the chemical potentials and the phase composition obtained from numerical simulation of the NRCH model. $f_{12} = \phi_1 \phi_2$, and the interaction parameters are indicated in the figure and $K_{12} = K_{21} = 0$.}
	\label{fig:Binodals}
\end{figure}

{\it{Conditions for phase coexistence--}} We first consider a binary system with densities $\phi_1$ and $\phi_2$. We redefine the coefficient as $\chi_{12} = \chi + \alpha$ and $\chi_{21} = \chi- \alpha$, using explicitly the coefficient of reciprocal and nonreciprocal interactions $\chi$ and $\alpha$. $\alpha$ is tuned at a value below the exceptional point, $|\alpha|<|\chi|$, such that the system exhibits phase separation. We begin by consider a quasi one-dimensional pattern, i.e. the density varies in only in a direction perpendicular to the flat interface which we choose to be the $x$ direction. 

We consider phase separation into two phases that we label $(1)$ and $(2)$. In phase 1, the composition is $( \phi_a^{\mo},\phi_b^{\mo} )$, while phase in 2 the composition is $(\phi_a^{\mt},\phi_b^{\mt} )$. When the mixture reaches a steady state that is invariant under time translation, the chemical potentials are constant throughout the system $\mu_a^{\mo} = \mu_a^{\mt} \equiv \langle \mu_a \rangle$ for both species -- taken together, these two conditions imposes chemical equilibrium. To find a condition similar to mechanical balance in this system out of equilibrium, we follow the procedure outlined in~\cite{GeneralisedThermo}. We proceed by multiplying $\mu_a$ by $\partial_x \phi_a$ and integrate from a point deep within one phase to another to obtain the following for species 1
\beq 
I_1 \equiv \int_{\mo}^{\mt} \mbox{d}x  \, \mu_1 \partial_x \phi_1 
 = \langle \mu_1 \rangle (\phi_1^{\mt} - \phi_1^{\mo}).
 \label{eq:eqIa}
\eeq
Using Eq. \eqref{eq:chemicalPotential} to evaluate $I_1$ we obtain an equivalent expression for the integral 
\beq
&& I_1 = \int_{\mo}^{\mt} \mbox{d}x  \,  \frac{\mbox{d}f_1}{\mbox{d}\phi_1}  \partial_x \phi_1 -  \int_{\mo}^{\mt} \mbox{d}x  K_{11} (\partial^2_x \phi_1)(\partial_x \phi_1)  \nonumber \\
&& + \int_{\mo}^{\mt} \mbox{d}x  \, [ (\chi+\alpha)   \frac{\partial f_{12}}{\partial \phi_1}  \partial_x \phi_1 -  K_{12}  (\partial^2_x \phi_2 )(\partial_x \phi_1)].
\label{eq:IntMu1}
\eeq 
The terms in the first line of Eq.~\eqref{eq:IntMu1} are total derivatives that depend only on the bulk values of the fields, in contrast to the terms in the second line with contributions from the interaction which are not. Defining a partial pressure like quantity $P_a^{(i)} = \langle \mu_a \rangle \phi_a^{(i)}  - f_a$ for species a in phase $(i)$, we can rewrite Eqs. \eqref{eq:IntMu1} and \eqref{eq:eqIa} as for species 1 as
\beq 
&&P_{1}^{\mt} - P_{1}^{\mo} =  (\chi + \alpha) \int_{\mo}^{\mt} \mbox{d}x  \,  \frac{\partial f_{12}}{\partial \phi_{1}}  \partial_x \phi_{1} \nonumber \\
&&-  \int_{\mo}^{\mt} \mbox{d}x  K_{12}  (\partial^2_x \phi_2)(\partial_x \phi_1).
\label{eq:partialPdiff}
\eeq
A similar relation holds for species 2 yielding an expression for $P_2^{\mt} - P_2^{\mo}$. Multiplying Eq.~\eqref{eq:partialPdiff} by $\chi-\alpha$ and the equivalent one for species 2 by $\chi+\alpha$, and adding up the contributions we obtain the following relation that relates a quantity on the L.H.S. that depends only on the phase composition with the quantity on the R.H.S. that depends on the detailed form of the interface in the steady state.  
\beq 
&& \left[ P_1 (\chi - \alpha)+ P_2 (\chi +\alpha) - (\chi^2 -\alpha^2) f_{ab}) \right]^{\mt}_{\mo} \nonumber \\
&& = -  \int_{\mo}^{\mt} \mbox{d}x  \left[ (\chi-\alpha) K_{12} (\partial^2_x \phi_2)(\partial_x \phi_1) \right.  \nonumber \\
&& \left. + K_{21} (\chi+\alpha)  (\partial^2_x \phi_1)(\partial_x \phi_2) \right]
\label{eq:InterfacialPressure}
\eeq 
The integrand on the R.H.S. of Eq.~\eqref{eq:InterfacialPressure} last line can be written as the a complete derivative of the function $\partial_x \phi_1 \partial_x \phi_2$, for $ (\chi-\alpha) K_{12} = K_{21} (\chi+\alpha)$. We identify the active pressure as the sum (difference) of two different terms
\beq
P &=&  P_{r} - \frac{\alpha}{\chi}P_{nr},
\label{eq:Pressure}
\eeq
where the reciprocal and the non-reciprocal contributions are as follows
\beq
P_r &=& P_1+P_2 - \chi f_{12} =  \mu_1 \phi_1 + \mu_2 \phi_2 - f, \nonumber \\
P_{nr} &=& P_2 - P_1 + \alpha f_{12}.
\label{eq:PressureParts}
\eeq 
The R.H.S. of Eq. \eqref{eq:InterfacialPressure} is the Laplace pressure difference across a flat interface - clearly a signature of an active system. Notice that the difference does not vanish when reciprocity is restored for the interfacial tension, that is for $K_{12} = K_{21}$. The expression in \eqref{eq:Pressure}, written such that $P$ reduces to the passive case when $\alpha = 0$, is invalid for $\chi = 0$. If interactions are completely non-reciprocal, the active pressure is $P = P_2 - P_1 + \alpha f_{12} = \mbox{constant}$, if $K_{12} = -K_{21}$. At the exceptional point $|\chi| = |\alpha|$, and for $\alpha>0$, the pressure reduces to the form $P=2 P_2$ and $P = 2P_1$ and $\alpha<0$.  This approach also yields the condition for phase coexistence for $K_{11} \neq K_{22}$, where we expect a conserved version of a Turing instability~\cite{Thiele_PhysRevE.103.042602}. We also find that the pressure reduces to the form derived in \cite{brauns2023nonreciprocal} when one of the species in the binary mixture does not phase separate in the absence of interactions.  

The limitation of the approach outlined for the binary system is that if we introduce another interaction term  $f'_{12}$ in Eq.~\eqref{eq:chemicalPotential}, with a functional form that is different from $f_{12}$ and using a different non-reciprocal coupling parameter $\alpha' \neq \alpha$, then we cannot use this method to obtain $P$. Strikingly, if the interactions are purely non-reciprocal, then the active pressure in presence of $f_{ab}'$ is given by a straightforward generalisation of Eq. \eqref{eq:Pressure} and \eqref{eq:PressureParts} - $P = P_1 - P_2 + \alpha f_{12}+ {\alpha}' f'_{12} $, with an effective free energy $f_1 - f_2 + \alpha f_{12}+ \alpha' f'_{12}$.

The existence of $P$ points at an underlying effective free energy. Consider the following for an effective bulk free energy density 
\begin{equation}\label{eq:EffF}
f = (1-\alpha \chi^{-1})f_1 + (1+\alpha \chi^{-1}) f_2 + \chi(1-\alpha^2 \chi^{-2}) f_{12}.
\end{equation}
Defining chemical potentials as derivatives of Eq.~\eqref{eq:EffF}, $\bar{\mu}_1 \equiv \delta f/\delta \phi_1 $, the pressure evaluated as $P = \bar{\mu}_1 \phi_1 + \bar{\mu}_2 \phi_2 - \bar{f}$, is identical to that in Eq. \eqref{eq:Pressure}. $\bar{\mu}_a$ is related to $\langle \mu_a \rangle$ as $\bar{\mu}_1 = \langle \mu_1 \rangle/(\chi+\alpha)$ and $\bar{\mu}_2 = \langle \mu_2 \rangle/(\chi-\alpha)$, and are well defined at all points other than the exceptional point $|\chi| = |\alpha|$. This implies that we can infer the phase composition using a common tangent construction on the effective free energy $\bar{f}$ at all values of $\alpha$ other than at $\chi$. We find that the free energy Eq.~\eqref{eq:EffF} has the same functional form as the Lyapunov function associated with the minimal oscillator defined as $\dot{x}_1 = -\mu_1({x_1,x_2})$. 

To illustrate the usefulness of \eqref{eq:Pressure} in determining the phase composition, we present a few concrete examples. We first consider the case $K_{12} = K_{21} = 0$. We choose simple polynomial forms for the free energy ${f}_a =  s_a {\phi_a^2}/{2} +  {\phi_a^4}/{4 c_a^2}$, where $s_a$ is constrained to be $\pm1$. For $s_a = -1$, the system partitions into two compartments with densities $\pm c_a$, which is chosen to be unity for all numerical simulations presented here. Equivalent to the passive case, we can use the equality of the active pressure and chemical potentials to determine the composition in the phases for binary or ternary phase separation, for details see SI. The results are shown in Fig. \ref{fig:Binodals}. For vanishing average density $\bar{\phi}_1 = \bar{\phi}_2 = 0$, the system partitions into two phases with compositions that differ in sign. The compositions lie in the minima of the potential $f$ in Eq.~\eqref{eq:EffF}. At the EP the $f$ is quasi one-dimensional such that $\mu_2=0$. At this point a saddle node bifurcation occurs where the minima splits into a saddle point and an unstable maxima. At large enough $\alpha$, $f$ is concave to the ($\phi_1 - \phi_2$) plane. We also choose a nonlinear form for the interactions $f_{12} = \phi_1 \phi_2(1+ \lambda \phi_1 \phi_2)$ to show the success of the common tangent composition in predicting the phase composition. 

For a vanishing $K_{ab}$, the non-reciprocal surface tension, the discontinuity in pressure disappears in one dimension and the mapping to the passive case is complete. The mapping to equilibrium phase separation fails when $K_{ab} \neq 0$ and the discontinuous jump in the pressure becomes relevant. We will also show later that interfacial effects in two are higher dimensions is subtle as the non-reciprocal interaction modifies the interfacial tension and thus the Laplace pressure within a droplet.
\begin{figure}
	\centering
	\includegraphics[width= 1\linewidth]{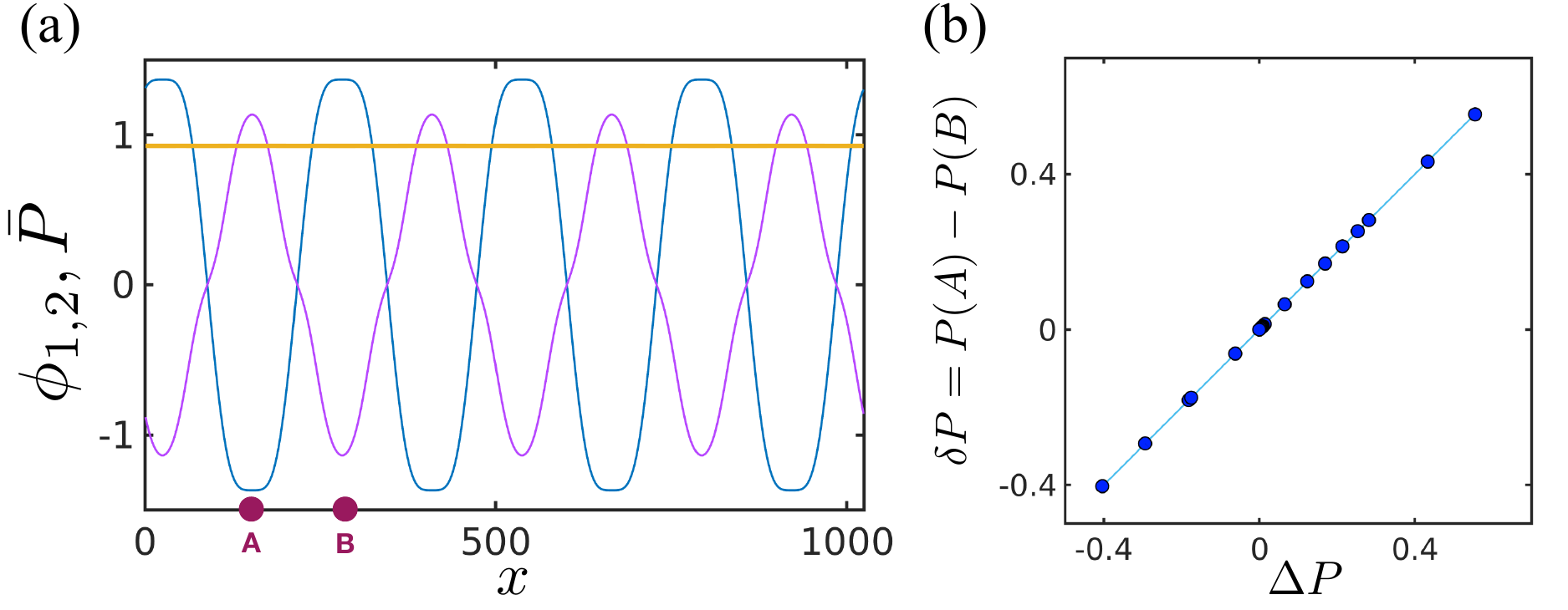}
\caption{{\textbf{Active Laplace pressure}} (a) Density profiles $\phi_1$ and $\phi_2$ in the steady state for $\chi$, $\alpha$ and $K_{12} = K_{21} = -5$ showing micro phase separation. (b) The active Laplace pressure is calculated by varying $K_{12}$ in the range $(-5,5)$. The active Laplace pressure difference between the pressure obtained from simulations and theory are plotted showing excellent agreement.}
	\label{fig:Pressure}
\end{figure}

{\it{Excess Laplace Pressure within a droplet--}}
To calculate the change in pressure across a curved interface, we consider a spherical droplet of size $R$. Integrating $\mu_a \partial_r \phi_a$ from $r =0$ to $r=\infty$, i.e. from inside a droplet to the outside we obtain the following expression for the excess pressure $\Delta P$ 
\beq 
&& \Delta P =    \nonumber \\
&& \int_0^{\infty} \mbox{d}r \left(1-\alpha \chi^{-1} \right) \left[    K_{11}   ( \phi^{\prime}_1)^2{r^{-1}}  + K_{12} {\phi'_1 \phi'_2 }{r^{-1}} \right]   \nonumber \\
&&  + \left(1+{\alpha}{\chi^{-1}} \right)  \left[ K_{22} {(\phi'_2)^2}{r^{-1}}  + K_{21} {\phi'_1 \phi'_2 }{r^{-1}} \right]   \nonumber \\
&&+  \left(1-{\alpha}{\chi^{-1}} \right) K_{12}  \phi^{\prime \prime}_1 \phi'_2   +  \left(1+{\alpha}{\chi^{-1}} \right) K_{21} \phi^{\prime \prime}_2 \phi^{\prime}_1 .
\label{eq:Laplace2D}
\eeq 
The first two lines on the R.H.S. of Eq. \eqref{eq:Laplace2D} leads to contributions that scale inversely with the droplet size $R$. At equilibrium, $\alpha = 0$ and $K_{ab} = K_{ba}$, $\Delta P$ is always a positive quantity, and it does not depend explicitly on the interaction parameter $\chi$. It receives contribution only from the first two lines which scale as $R^{-1}$.  For nonzero $\alpha/\chi$ appears explicitly in the expression for $\Delta P$.  For non-reciprocal surface tension, $\Delta P$ can be a negative thus signalling a clear departure from equilibrium. 
\begin{figure}
	\centering
	\includegraphics[width= 1 \linewidth]{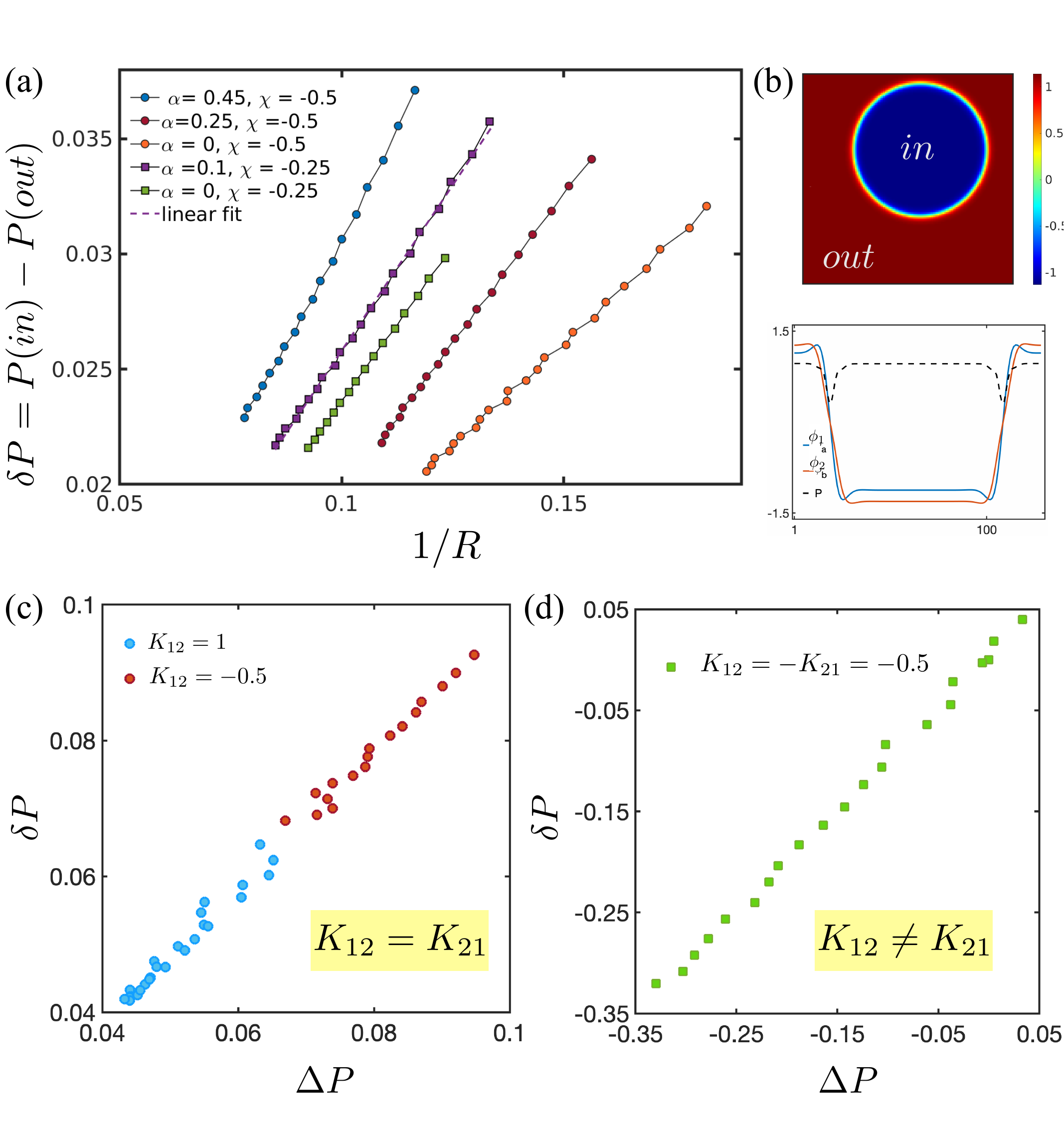}
	\caption{{\textbf{Active Laplace pressure in a droplet--}} (a) For $K_{12} = K_{21} = 0$, the excess pressure $\Delta P$ inside a droplet scales linearly with inverse of the droplet size as it does in passive systems. Panel (a) shows a few instance of the excess Laplace pressure $R^{-1}$, see the interaction parameters in the inset and the parameters in the Appendix. (b) A heatmap showing a droplet at steady state for $K_{12} = K_{21} = 5$. The density profiles across a line intersecting the centre of the droplet are shown below the heatmap. (c) Laplace pressure for $K_{12} = K_{21}$ evaluated for values indicated. The excess pressure $\Delta P$ as obtained from simulations is plotted against the analytical expression in Eq. \eqref{eq:Laplace2D} showing good agreement.  (d) Same for $K_{12} = K_{21} = -0.5$, i.e. fully anti-symmetric interfacial tension coefficients, showing that the theory fully captures the active Laplace presssure measured in simulations. }
	\label{fig:Pressure}
\end{figure}

{\it{Pressure in a multicomponent system--}}
We now show that we can find conditions for phase equilibrium for an arbitrary number of components following the steps outlined for $N=2$. Consider, same as was done previously for the binary system, that the system partitions into two compartments with compositions labelled 1 or 2. By summing over all species $b$ we can write an expression similar to Eq. \eqref{eq:IntMu1} 
\beq 
P_{a}^{(2)} - P_a^{(1)} =  \sum_{b \neq a} \chi_{ab} \int_{\mo}^{\mt} \mbox{d}x  \,  \frac{\partial f_{ab}}{\partial \phi_a}  \partial_x \phi_a
\label{eq:PressureGen}
\eeq 
Multiplying both sides of Eq.~\eqref{eq:PressureGen} by a constant $\lambda_a$ and summing over all components we want to write Eq. \eqref{eq:PressureGen} in the following form 
\beq 
P^{(2)} - P^{(1)} \equiv \left[\sum_a \lambda_a P_a  - \sum_{\langle ab \rangle} \lambda_a \chi_{ab} f_{ab}\right]^{\mt}_{\mo} = 0,
\label{eq:GeneralCon}
\eeq
where $<ab>$ denotes distinct pairs, and thus obtain an expression for the pressure $P$. In other words, the goal is to find a set of coefficients $\{\lambda_a\}$ such that the integrals on the R.H.S. of Eq.~\eqref{eq:PressureGen} can be directly evaluated and Eq. \eqref{eq:GeneralCon} is satisfied. This is possible strictly if, given any pair of species $a$ and $b$ the following relation holds for each distinct pair
\beq
\frac{\lambda_a}{\lambda_b} = \frac{\chi_{ba}}{\chi_{ab}}. 
\label{eq:Ratio}
\eeq 
Eq. \eqref{eq:Ratio} represents $N(N-1)/2$ equations that are used to determine $N-1$ free coefficients $\{ \lambda_a \}$. The remaining $(N-1)(N-2)/2$ equations constrain the interaction coefficients $\chi_{ab}$. A system with N components can thus be mapped to the equilibrium case only in the restricted case where the $N(N-1)$ parameters of the interaction matrix are constrained by $(N-1)(N-2)/2$ equations, i.e. $(N+2)(N-1)/2$ of the parameters are free thus implying that the mapping to equilibrium is possible only for a restricted set of interaction coefficients $\chi_{ab}$.

\begin{figure}
	\centering
	\includegraphics[width= 0.8\linewidth]{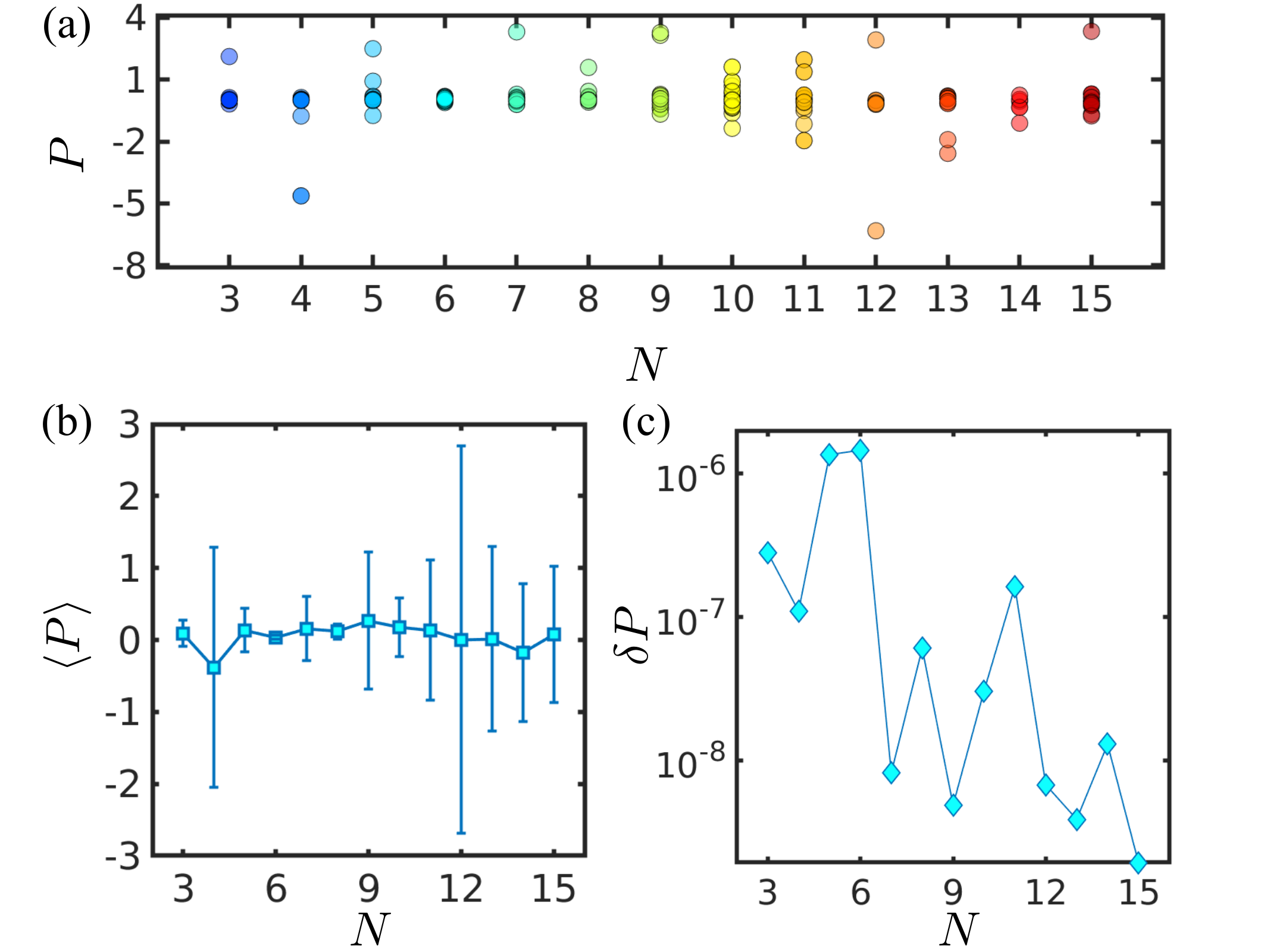}
	\caption{{\textbf{Pressure in multicomponent systems--}} (a) The validity of the concept of pressure is checked numerically in multi-component systems. The markers denote $P$ calculated in the steady state for $10$ examples for all values of $N$. (b) Shows the average value of $P$ as a function of $N$. (c) Spatial deviation from a constant $P$ averaged over the $10$ realisations shows that the deviations from a constant P is negligible. }
	\label{fig:BinodalsMC}
\end{figure}

For illustration we consider $N=3$ for which the interaction matrix is characterised by 6 parameters. We will show that for a mapping to equilibrium, we need to constrain the parameters with a single constraint consistent with the conclusions for the general $N$. The three equations for $\lambda_a$ are
\beq 
\frac{\lambda_1}{\lambda_2} = \frac{\chi_{21}}{\chi_{12}}, \,\frac{\lambda_1}{\lambda_3} = \frac{\chi_{31}}{\chi_{13}},\, 
\frac{\lambda_2}{\lambda_3} = \frac{\chi_{32}}{\chi_{23}}.
\label{eq:ratio3}
\eeq 
Choosing $\lambda_1 = 1$, we have $\lambda_2 = {\chi_{12}}/{\chi_{21}}$ and $\lambda_3 = {\chi_{13}}/{\chi_{31}}$ from the first two equations in \eqref{eq:ratio3}. The third equation constraints the coefficients as
\beq 
\chi_{12} \chi_{31} \chi_{23} \chi_{21}^{-1}\chi_{13}^{-1} \chi_{32}^{-1} = 1. 
\label{eq:compNcoeff2}
\eeq
To verify Eq. \eqref{eq:GeneralCon}, we have run simulations in one dimension varying the total number of components from $3-15$ and for about 10 instances of the interaction $\chi_{ab}$ and verify that at steady state, $P$ is indeed constant throughout the system for all components, see Fig.~\ref{fig:BinodalsMC} (a) with negligible variance as shown in Fig.~\ref{fig:BinodalsMC} (b-c). To show this we choose $K_{aa}=1$ for all $N$ species and all cross-couplings $K_{ab}=0$. The results are shown in Fig. \ref{fig:BinodalsMC}

{\it{Conclusion--}}
To conclude, we have obtained an expression for pressure in nonreciprocally interacting scalar densities which can be used to determine phase coexistence in a broad class of systems with steady states comprising both bulk phase separated and micro-phase separated states. Counter-intuitively, a system with exclusively inter species nonreciprocal interactions can always be mapped to a free energy, phase separation occurs if the free energy allows a convex hull construction. The concepts explored in this paper can be extended to other forms of active mixtures with orientational degrees of freedom~\cite{Fruchart2021,BenoitYu2023}.  

Pressure plays a more fundamental role than being just a condition determining phase coexistence in equilibrium physics. A question that arises naturally -- consider a collection of diffusiophoretic colloids~\cite{Ramin-Liverpool-Ajdari2005} which generically have nonreciprocal interactions between them and confine them to an enclosure, can we relate the average force per unit area to the expression obtained here? That is, does the pressure described here truly enforce mechanical equilibrium?

\section{acknowledment}
SS thanks B Mahault for multiple insights and recommendations, R Golestanian for many discussions and encouragement, and the HPC facility and maintenance team at LMP, MPIDS for their help and support. SS thanks V Novak for help with the graphics.

\end{document}